\documentclass[pra,twocolumn,preprintnumbers,amsmath,amssymb,superscriptaddress,nolongbibliography]{revtex4-1}
\usepackage{graphicx}
\usepackage{dcolumn}
\usepackage{bm}

\usepackage{color}
\usepackage[normalem]{ulem}
\usepackage{hyperref}

\begin{document}

\title{Crystal nucleation in a vapor deposited Lennard-Jones mixture}

\author{Fabio Leoni}
\affiliation{Dipartimento di Fisica, Universit\`a degli Studi di Roma La Sapienza, Piazzale Aldo Moro 5, Rome, 00185, Italy}
\author{Hajime Tanaka}
\affiliation{Department of Fundamental Engineering, Institute of Industrial Science, The University of Tokyo, 4-6-1 Komaba, Meguro-ku, Tokyo, 153-8505, Japan}
\affiliation{Research Center for Advanced Science and Technology, The University of Tokyo, 4-6-1 Komaba, Meguro-ku, Tokyo, 153-8904, Japan}
\author{John Russo}\email{corresponding author: john.russo@uniroma1.it}
\affiliation{Dipartimento di Fisica, Universit\`a degli Studi di Roma La Sapienza, Piazzale Aldo Moro 5, Rome, 00185, Italy}

\begin{abstract}
Understanding the pathways to crystallization during the deposition of a vapor phase on a cold solid substrate is of great interest in industry, e.g., for the realization of electronic devices made of crystallites-free glassy materials, as well as in the atmospheric science in relation to ice nucleation and growth in clouds.
Here we numerically investigate the nucleation process during the deposition of a glassformer by using a Lennard-Jones mixture, and compare the properties of this nucleation process with both its quenched counterpart and the bulk system.
We find that all three systems homogeneously nucleate crystals in a narrow range of temperatures.  
However, the deposited layer shows a peculiar formation of ordered domains, promoted by the faster relaxation dynamics toward the free surface even in an as-deposited state. In contrast, the formation of such domains in the other systems occurs only when the structures are fully relaxed by quenching. Furthermore, the nucleus initially grows in an isotropic symmetrical manner, but eventually shows sub-3D growth due to its preference to grow along the basal plane, irrespective of the layer production procedure.
\end{abstract}


\maketitle

\section{Introduction}
\label{sec:introduction}

The pathways to crystallization during the deposition of a vapor phase on a cold solid substrate are of great interest in a variety of contexts. For example, in the tech industry, it is crucial to realize electronic devices made of crystallites-free glassy materials, such as Organic Light-Emitting Diode (OLED) displays \cite{Ediger2017,Rodriguez2022}. 
What makes glasses ideal materials for such applications is their remarkable spatial homogeneity on a macroscopic scale free from localized defects. This unique property holds true for both single-component and  multi-component systems, as long as the components have the ability to mix.
On the other hand, polycrystalline materials can be very sensitive to the presence of defects such as grain boundaries which cause diffraction and then the loss of electric signals \cite{Ediger2017}. 
In atmospheric science, it is debated whether ice nucleation in clouds occurs directly via vapor deposition on ice nucleating particles present in the atmosphere, or if it occurs homogeneously from liquid water that condenses in cavities found on the surfaces of ice nucleating particles due to the ``inverse Kelvin'' effect~\cite{kanji2017}.
For a repulsive colloidal model, it has been observed that the surface confining the liquid may favor the heterogeneous or the homogeneous nucleation depending on its structure \cite{espinosa2019}.  
Water as a polymorphic and polyamorphic substance can show more complex crystallization mechanisms, both in bulk \cite{leoni2021}, in droplets \cite{espinosa2018}, and in vapor deposition systems \cite{tonauer2023}, with respect to liquids that admit one crystal phase. For example, simulations of vapor deposition of the monatomic water model mW on graphitic surfaces have shown that with increasing the temperature of the substrate, bilayer ice, ice I, and liquid droplets are nucleated in the deposited amorphous ice layer \cite{lupi2014}.

In the last 15 years, vapor deposition has been employed in experiments, as well as more recently studied with simulations, to obtain so-called ultra-stable glasses (USG) thanks to the fine-tuning of the substrate temperature.
USG shows extraordinary kinetic and thermodynamic stability. Achieving such stability in an ordinary glass would require significantly longer timescales, ranging from hundreds to even hundreds of thousands of years~\cite{Ediger2017}.
Simulations of vapor deposition mimicking experiments \cite{malshe2011,Lyubimov2013,Berthier2017,reid2016age,moore2019}
have shown that, in a range of substrate temperatures, the enhanced mobility at the free surface of the deposited layer is responsible for its ultra-stability. In particular, this mechanism has been recently shown \cite{leoni2023} to be crucial also for the stability of the glass-former model investigated in the present work (see below).
For stabilizing an ultrastable glass, several possibilities have been explored, such as having locally favored structures incompatible with the crystal symmetry~\cite{Shintani2006}, a large crystal unit cell extending beyond locally favored structures~\cite{Pedersen2010,pedersen2021}, competing interactions~\cite{Russo2018}, and/or compositional frustration~\cite{Hu2020}.

A recent study~\cite{leoni2023} has shown that a deposited glass composed of an equimolar additive bi-disperse Lennard-Jones (LJ) mixture, known as Wahnstr\"om (WAHN) model~\cite{Wahnstrom1991}, outperforms in stability the conventional glass version of the same material when the substrate is kept at a temperature in a specific range of values below the glass transition temperature $T_g$. Furthermore, it is shown that the enhanced mobility at the free surface of the deposited layer triggers the formation of locally favored structures (LFS) \cite{leoni2023} (see below), which are the signature of an ultrastable glass with low free energy. At the same time, LFS driven by local free-energy minimization promotes crystallization if their structures are compatible with a crystal~\cite{Tanaka2012,russo2016crystal}. Thus, it is an intriguing question how this apparently conflicting features hidden in LFS affect the stability of a ultrastable glass.

The WAHN system is an ideal model system to address this question since its LFS's, icosahedra, are also found in the crystal phase, which has been found in simulations to be a Frank-Kasper crystal ~\cite{Pedersen2010}. This model offers both the possibility to study the USG configurations below $T_g$, where LFS are correlated with the stability of the glassy phase (see Ref.~\cite{leoni2023}) then disfavoring crystallization, and to perform a straightforward characterization of the solid phase during the nucleation of the crystal phase above $T_g$, where, after a long relaxation, the appearance of Frank-Kasper crystal structures involves the formation of LFS (an aspect to be taken into account when applying the present results on nucleation to other glassformers).
Moreover, it is suitable to describe metallic glasses, which are of great interest for applications due to their unique mechanical properties \cite{wang2009,Ashby2006}.

\section{Methods}
\label{sec:methods}

Here we focus on the nucleation of the WAHN model by performing molecular dynamics simulations (using LAMMPS \cite{LAMMPS}) of vapor deposition, the quenched solid, and the bulk.
The WAHN model is described by the potential
\begin{equation}
u_{LJ}(r)=4\epsilon_{\alpha\beta}\left[\left(\dfrac{\sigma_{\alpha\beta}}{r}\right)^{12}-\left(\dfrac{\sigma_{\alpha\beta}}{r}\right)^6\right],
\end{equation}
where $\alpha$ and $\beta$ can be particles of type 1 and 2. The energies are $\epsilon_{11}=\epsilon_{12}=\epsilon_{22}=\epsilon$, the masses $m_2/m_1=2$, and the diameters $\sigma_{22}/\sigma_{11}=1.2$ with the cross-interaction diameter given by the Lorentz-Berthelot mixing rule: $\sigma_{12}=(\sigma_{11}+\sigma_{22})/2$.
We set $\sigma_{11}=\sigma$.
The units of energy, length and time are $\epsilon$, $\sigma$ and $t^*=(m_1\sigma^2/\epsilon)^{1/2}$, respectively. Quantities are expressed in these reduced units. The chosen integration time step is 0.005. 

Similarly to Refs.~\cite{Lyubimov2013,Berthier2017,leoni2023}, 2000 particles of type 1 and 2000 particles of type 2 are introduced in the box alternately from a random position in the xy plane at the top of the simulation box with a fixed vertical velocity $v_z=0.1$ and lateral components $v_{x,y}$ randomly extracted between -0.01 and 0.01. 
The frequency with which particles are introduced in the system is described by the deposition rate $\gamma_{DL}$, given by the ratio between the thickness of the deposited layer and the time elapsed to make it: $\gamma_{DL}=\Delta z/\Delta t$. We inject one particle every $5\cdot 10^4$ integration steps, so that we obtain $\gamma_{DL}\simeq 10^{-7}\sigma/dt=2\cdot 10^{-5}$.
As a reference, to convert internal units in real numbers, we consider type 1 particles as Argon atoms with $\epsilon/k_B=120$~K, where $k_B$ is the Boltzmann constant, and $t^*=2.2$~ps \cite{Pedersen2010}, such that $\gamma_{DL}\sim 1$~K/ns.

\begin{figure}[!t] 
\begin{center}
\includegraphics[width=4.2cm]{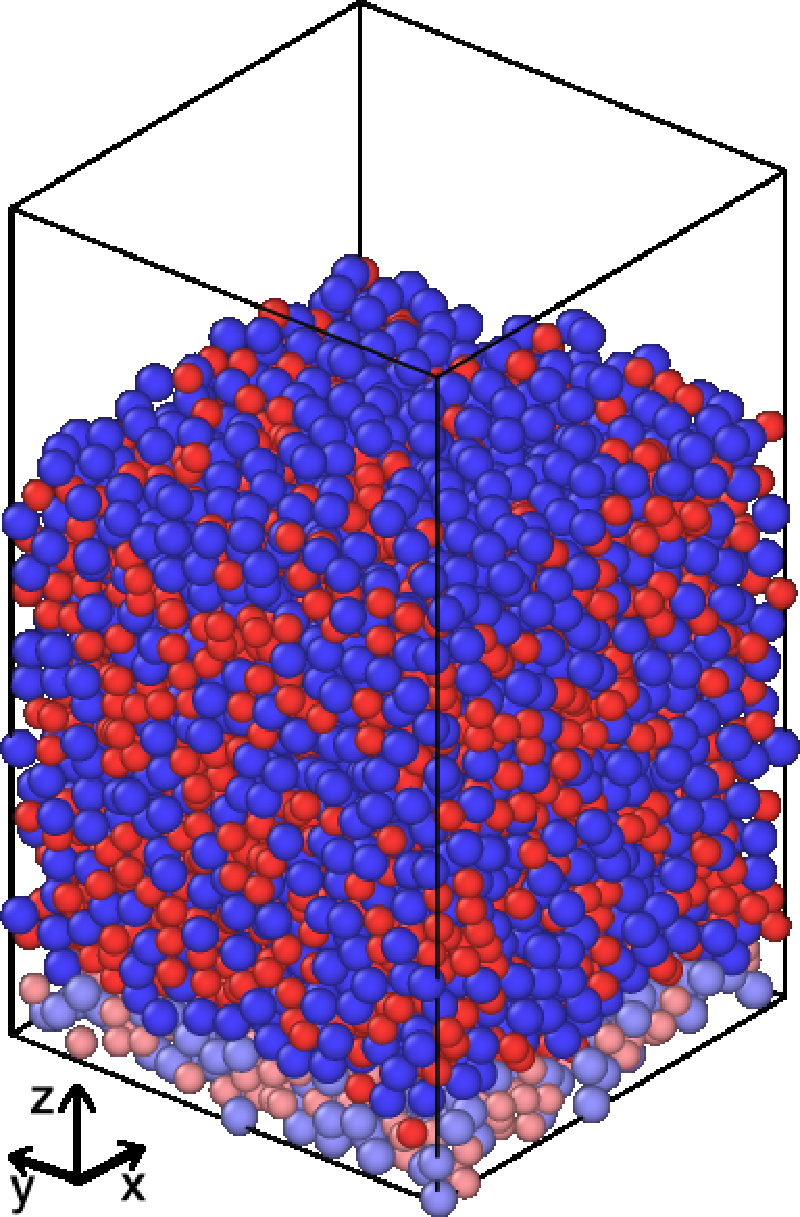}
\includegraphics[width=4.2cm]{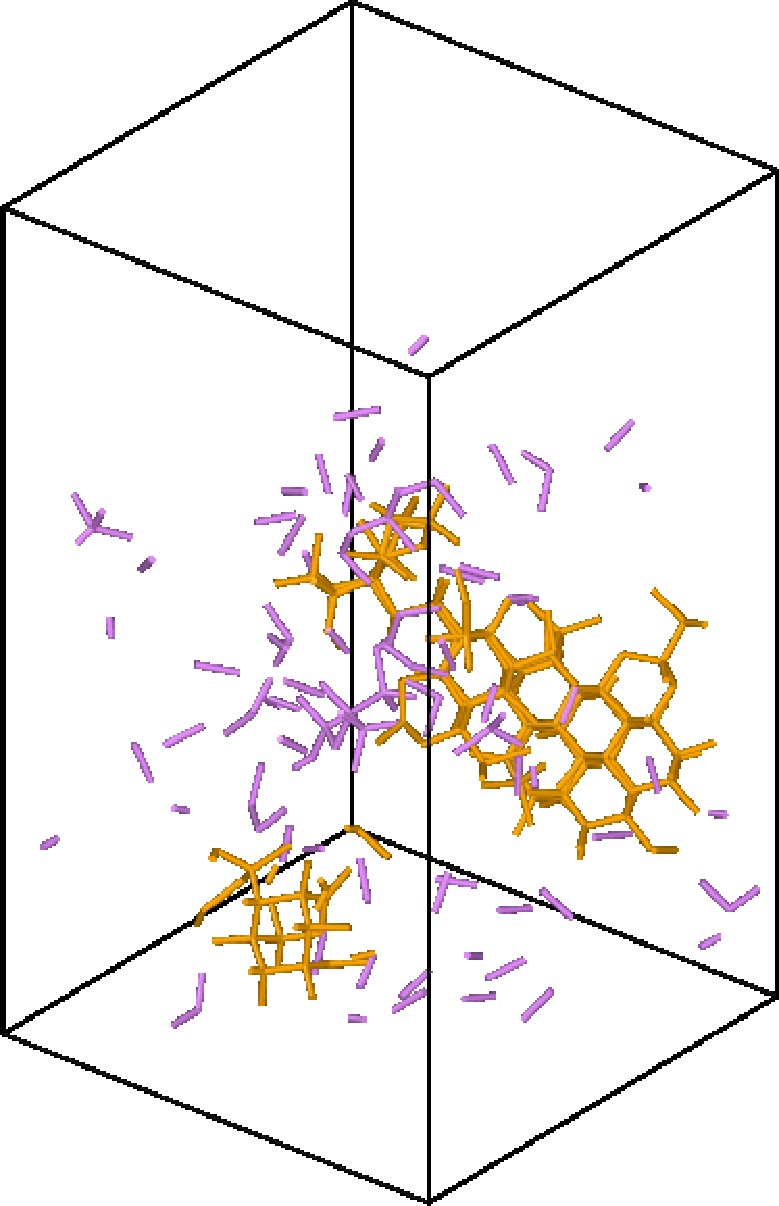}
\caption{\label{fig:snapshots} Left: snapshot showing for $T=0.4$ the as-deposited layer formed by particles of type 1 (red) and 2 (blue) and the substrate formed by particles of type 1 (light red) and 2 (light blue).
Right: snapshot showing Frank-Kasper bonds for the same configuration with particles of type 2 belonging to the main cluster connected by orange FK bonds and belonging to smaller clusters by violet FK bonds. 
The snapshots are made with Ovito \cite{Stukowski2009}.}
\end{center}
\end{figure}
The top edge of the box elastically reflects particles that, eventually, at high T, bounce or detach from the free surface.  
Injected particles are integrated in the NVE ensemble.
The substrate placed at the bottom of the simulation box (see Fig.~\ref{fig:snapshots}) is composed of a disordered WAHN mixture with 250 particles of type 1 and 250 of type 2 at the number density in reduced units $\rho^*=\rho(\sigma)^3=0.75$, and kept at temperature $T$ by using the Nos\'e-Hoover thermostat (NVT ensemble).
The configuration of the substrate is obtained by cutting a bulk system equilibrated at a temperature $T=0.4$ (in units of $\epsilon/k_B$) and a number density in reduced units which is $\rho^*=\rho(\sigma)^3=0.75$.   
The choice to take the substrate made of the same particles deposited on it avoids the possible formation of a gap between the deposited layer, and the substrate \cite{Lyubimov2013}.
The lateral size of the box is $L_x=L_y=17.5\sigma$.
A spring force of stiffness $k=50$ is independently applied to each substrate particle to tether it to its initial position. This value of $k$ is large enough to keep particles fluctuating around their initial position and small enough to allow the equilibration of the deposited layer. As we verified by simulations, below the glass transition temperature at zero pressure, $T_g\simeq0.36$ \cite{leoni2023}, the use of springs is not necessary to keep the particles of the substrate confined at the bottom of the box (in which case the bottom edge of the box is reflective as the top edge), while above $T_g$ they prevent these particles from mixing with the deposited particles. 

To investigate the effect of the deposition process on nucleation, we compare the deposited glass layer (DL) with a quenched glass layer (QL) whose quenching rate is comparable to DL (see below) and the bulk glass.
The QL is obtained by melting the DL (by instantaneously heating the system temperature to $T=1$), letting it equilibrate for a time of $5\cdot 10^3$ (in reduced units), and then cooling it down to $T=0.1$ in a time equal to $5\cdot 10^4$, so that the resulting cooling rate $\gamma_{QL}=1.8\cdot 10^{-5}$ is comparable to $\gamma_{DL}$.
During the cooling procedure, we keep the substrate in contact with the deposited layer with springs active during heating and cooling.
In this case, to make the QL, the full system (deposited layer plus substrate) is in contact with a thermal bath (i.e., $T$ refers to the full system).
Since the presence of a free surface implies a vapor pressure, we consider the bulk case at pressure $P=0$. The bulk system's size is 4000 particles, the same as the DL and QL.
In the following section, we compare the DL and QL with their relaxed versions obtained by putting the full system in contact with a thermal bath.

\begin{figure}[t]
\begin{center}
\includegraphics[width=8.5cm]{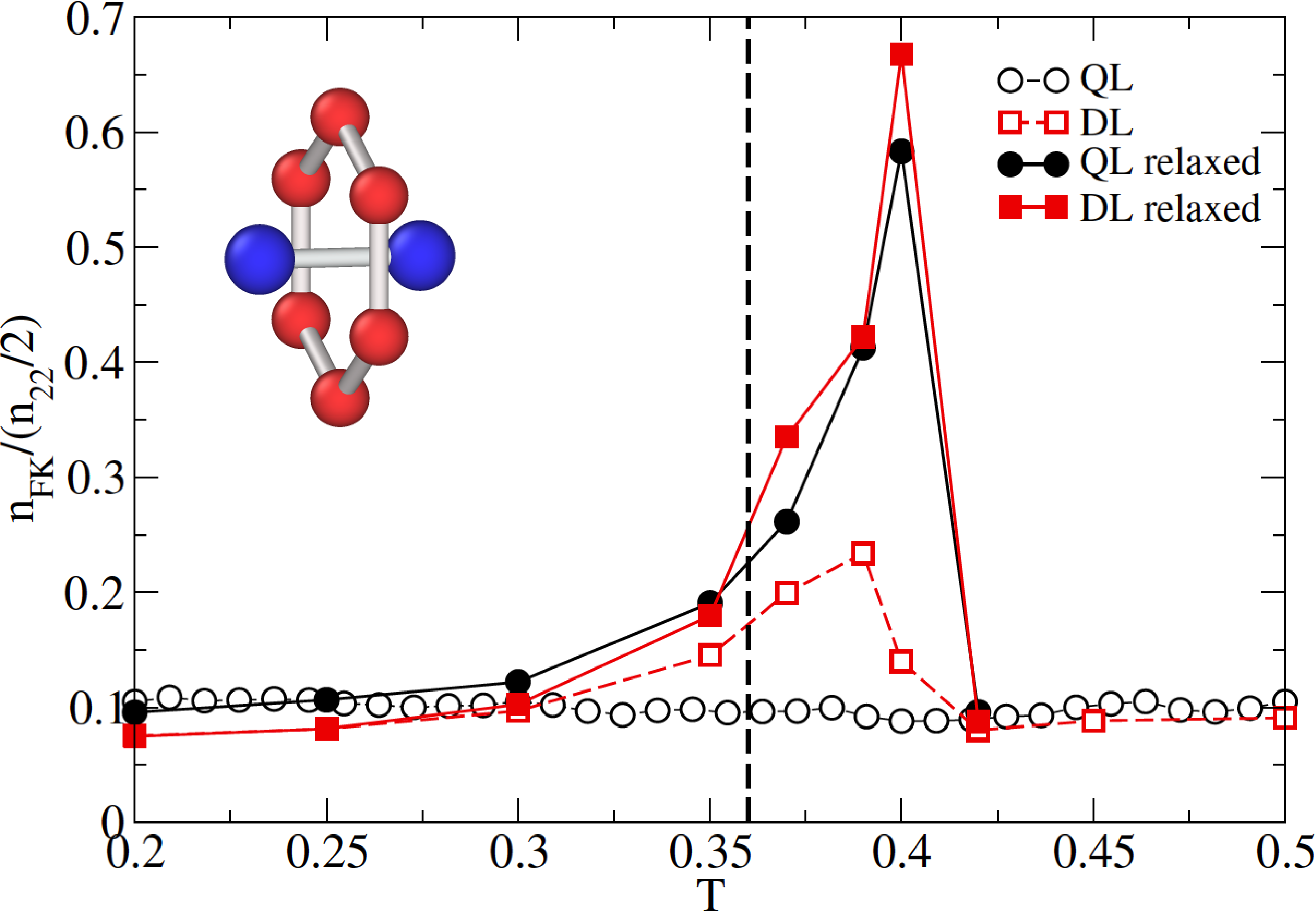}
\caption{\label{fig:FK} Number of Frank-Kasper bonds, $n_{FK}$,  normalized for half the number of particles of type 2, $n_{22}$, versus temperature $T$. DL as-deposited (open red squares) and relaxed (full red squares) and QL (open black circles) and relaxed (full black circles). The snapshot in the inset shows a schematic FK bond between two type 2 particles (in blue).}
\end{center}
\end{figure}

WAHN mixtures eventually crystallize into a structure of composition A$_2$B \cite{Pedersen2010}, where A and B correspond to particles of type 1 and 2, respectively. Crystals with this structure, such as MgZn$_2$, MgCu$_2$ and MgNi$_2$, are known as Laves phases \cite{Pedersen2010}.
To detect the formation of crystalline nuclei in our simulations, we look for the presence of
Frank-Kasper (FK) polyhedra \cite{frank1958} adapted to WAHN mixtures \cite{Pedersen2010}.
FK bonds connect two neighboring particles of type 2 if they share at least 6 neighbors of type 1 (see a snapshot in Fig.~\ref{fig:FK}). We compute neighbors using the radical Voronoi algorithm suited for additive mixtures \cite{rycroft2009}. The crystalline nuclei are composed of a tetrahedral network of FK bonds. While in previous works \cite{Pedersen2010,pedersen2021} only type 2 particles from which 4 FK bonds emanate towards other neighboring type 2 particles are considered to belong to the crystal phase, here we identify crystal clusters as composed of neighboring type 2 particles linked by an FK bond. We found that the shape of the nuclei is not affected by this definition.

\section{Results and discussions}
\label{sec:results}

After relaxing configurations of as-deposited layers up to a maximum of $10^9$ time steps (i.e., $5\cdot 10^6$ in internal units), we find crystallization in the narrow range of temperatures $0.395\le T\le 0.41$.
In Fig.~\ref{fig:FK}, we show the number of FK bonds $n_{FK}$ normalized for half the number of particles of type 2, $n_{22}$, versus $T$ for both the deposited and supported layers.
The as-deposited layer (open red squares) shows a peak of FK bonds around $T=0.39$, while the supported layer (open black circles) shows a flat profile.
After relaxation, the peak of the DL (full red squares) increases in intensity and shifts to $T=0.4$, and also the QL (full black circles) develops a similar peak.

\begin{figure}[t]
\begin{center}
\includegraphics[width=8.5cm]{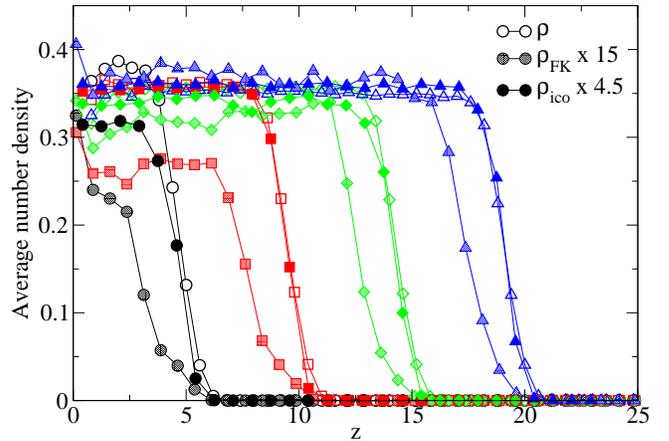}
\caption{\label{fig:comparison_all_rho_T0.42} Average number density profile along $z$ for particles distribution (open symbols), FK bonds multiplied by a factor $15$ (partially-filled symbols) and icosahedra multiplied by a factor $4.5$ (full symbols) at 4 different times during deposition ($t_1$, $t_2$, $t_3$, $t_4$ corresponding to black circles, red squares, green diamonds, and blue triangles, respectively) for $T=0.42$.}
\end{center}
\end{figure}
First, we investigate the structure of the layer at a temperature ($T=0.42$) where it reaches equilibrium and no nucleation occurs within the maximum time simulated. 
In addition to particle number density and FK bonds we characterize local order through the locally favored structures (LFS) defined as those geometric motifs which allow to identify molecular arrangements corresponding to some local minima of the free energy \cite{Jenkinson2017}. 
LFS in the Wanstr\"om model has been shown to consist of particles forming icosahedra \cite{malins2013identification,Wahnstrom1991,Coslovich2007}. 
In \cite{leoni2023} it has been found in WAHN a strong correlation between the number of LFS and temperature, a result also found in systems as different as colloids \cite{Royall2008b,Leocmach2012}, metallic glasses \cite{Royall2015}, or water in combination with a neural network classification scheme \cite{Martelli2020}. 
We identify icosahedra with the topological cluster classification (TCC) algorithm \cite{Malins2013}.

\begin{figure}[t]
\begin{center}
\includegraphics[width=8.5cm]{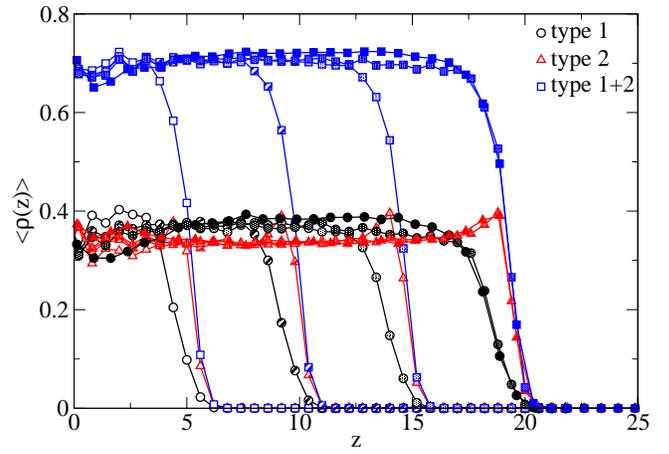}
\caption{\label{fig:rho} Average number density profile $\rho(z)$ of type $1$, $2$ and $1+2$ particles along $z$ for $T=0.4$ at 4 different times during deposition ($t_1$, $t_2$, $t_3$, $t_4$ corresponding to symbols going from open to fully filled, respectively).}
\end{center}
\end{figure}

We compare the average number density profile along $z$ for particles distribution ($\rho$), for FK bonds ($\rho_{FK}$) and for icosahedra ($\rho_{ico}$) at $T=0.42$ after relaxing each one of the four systems obtained by depositing particles up to a time $t_1$, $t_2$, $t_3$ and $t_4$ ($t_1=t_d/4$, $t_2=t_d/2$, $t_3=3t_d/4$, $t_4=t_d$, where $t_d=10^6$ is the time duration of the full deposition process), i.e., by depositing a fraction $1/4, 2/4, 3/4$, and $1$ of the full DL system. 
Each system is relaxed for a time equal to $t_r=2.5\cdot 10^6$, and averages are computed considering only the second half of the relaxation process for a single realization of the substrate.
In order to compare $\rho_{FK}$ and $\rho_{ico}$ with $\rho$ on the same scale, both densities are rescaled by a factor 15 and 4.5, respectively.
Figure~\ref{fig:comparison_all_rho_T0.42} shows that the distribution of icosahedra mirrors the distribution of particles at liquid-vapor interfaces as expected~\cite{godonoga2011}, while FK bonds are suppressed near the free surface by about an extra $2\sigma$ length due to the wider range involved in the formation of an FK bond compared to an icosahedron, and to the segregation of type 2 particles near the free surface (see below and Fig.~\ref{fig:rho}).
\begin{figure}[t]
\begin{center}
\includegraphics[width=4.6cm]{Fig_4a.eps}
\includegraphics[width=3.9cm]{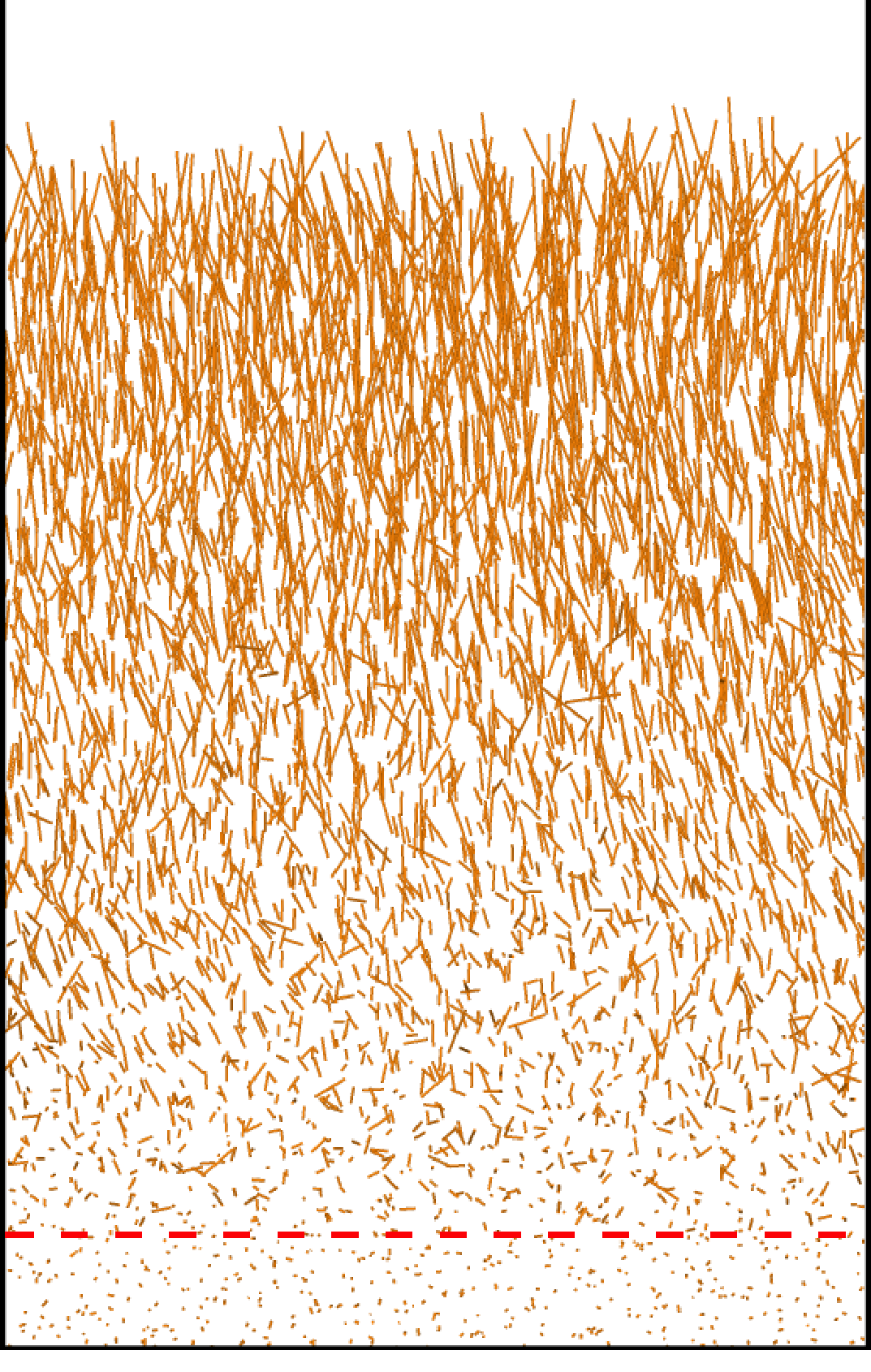}
\caption{\label{fig:FIRE} Left: position $z$ versus average displacement along $z$ ($\delta_{\perp}$) of type 1 (black circles) and type 2 (red squares) particles, and along a direction in the xy plane ($\delta_{\|}$) of type 1 (green circles) and type 2 (blue squares) particles after FIRE minimization applied to the as-deposited glass at $T=0.42$.
Right: snapshot showing the displacement vector from the as-deposited position of each particle and the respective position after minimization. The dashed red line shows the location of the dividing surface between the substrate and the deposited layer.}
\end{center}
\end{figure}

To study the stability of the DL at $T=0.42$, we perform an energy minimization of the as-deposited system by using the FIRE algorithm \cite{bitzek2006}, which employs a damped dynamics method to adjust particle coordinates giving the inherent structures (IS). 
In the right panel of Fig.~\ref{fig:FIRE}, we show a snapshot of the displacement vectors ${\bf r}_{IS}-{\bf r}_{0}$ for each particle in the system, where ${\bf r}_0$ and ${\bf r}_{IS}$ are the position before and after minimization, respectively. 
In the left panel of Fig.~\ref{fig:FIRE} we show the average displacement along the $z$ direction, $\delta_{\perp}=\langle |z_{IS}-z_0|\rangle$, and along a direction within the $xy$ plane, $\delta_{\|}=\langle (|x_{IS}-x_0|+|y_{IS}-y_0|)/2\rangle$ for particle type 1 and 2 separately.
Figure~\ref{fig:FIRE} shows that for both type 1 and 2 particles $\delta_{\perp}$ increases with increasing $z$, that is, when approaching the free surface, while $\delta_{\|}$ remains constant. This behavior is reminiscent of that of a free-standing layer \cite{Shi2011} made of Kob-Andersen (KA) \cite{Kob1994} particles, where energy minimization displaces particles near both free surfaces towards the center of the free-standing layer.
Interestingly, this behavior persists even in the presence of a single free surface, as shown in Fig.~\ref{fig:FIRE}. These findings further support the notion that free surfaces play a crucial role in enhancing the efficiency of landscape sampling.

To investigate the nucleation process during deposition, we perform simulations at $T=0.4$ for 30 different realizations of the system (different spatial disordered configurations of the substrate keeping fixed its density to $\rho^*=\rho(\sigma)^3=0.75$, and different random sequences of initial position and lateral velocities of injected particles).
We compute properties of the deposited layer at four times during deposition ($t_1$, $t_2$, $t_3$, and $t_4$) and at $t_r$ and $2t_r$ for the relaxed layer.

First, we compute the number density profile along the direction $z$ for type 1, 2 and 1+2 particles (see Fig.~\ref{fig:rho}).
We notice that the density profile in the middle of the pore is always flat and sharply drops to zero near the free surface (within $\sim 2\sigma$). Particles of type 2 (of larger diameter) show segregation at the free surface. 

This effect has been observed in previous simulations for the deposition of KA mixtures in 2D \cite{reid2016age} and 3D free-standing layers of KA \cite{Shi2011}. In these works, this effect has been attributed to the maximization of type 1 -- type 2 interactions in the layer since for KA mixtures (where type 1 and 2 are commonly indicated with B and A particles, respectively) $\epsilon_{12}>\epsilon_{22}>\epsilon_{11}$ and type 2 and 1 particles are $80\%$ and $20\%$ of the total, respectively. 
In the present case, type 2 and 1 particles are present in the same proportion, and all the $\epsilon$ are the same. Therefore, the segregation of type 2 particles we observe at the free surface should be linked to the system's minimization of the energetic cost associated with interface formation through the maximization of the number, rather than the type, of particle interactions, coming from optimal particle packing. This effect could be relevant in explaining segregation phenomena in the Kob-Andersen (KA) model as well.

\begin{figure}[t]
\begin{center}
\includegraphics[width=8.5cm]{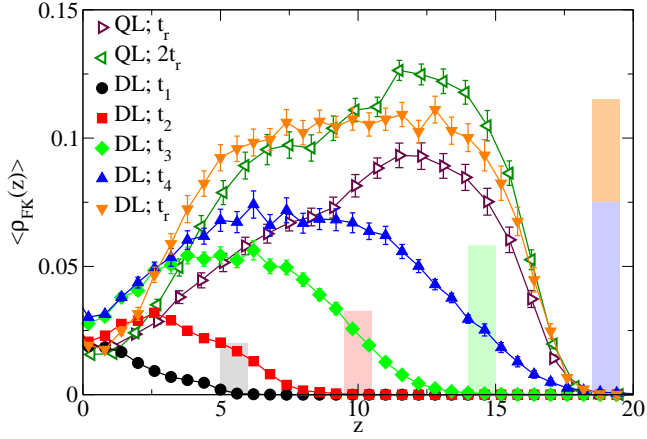}
\caption{\label{fig:rho_FK} Average FK bonds density profile $\rho_{FK}(z)$ along $z$ for $T=0.4$ at 4 different times during deposition ($t_1$, $t_2$, $t_3$, $t_4$ corresponding to full black circles, red squares, green diamonds, and blue upward triangles, respectively) and after a relaxation time $t_r$ (downward magenta triangles). Vertical bands show the average position of the free surface at times $t_1$, $t_2$, $t_3$, $t_4$ following the same color code of the DL.}
\end{center}
\end{figure}

After we verified that the density profile along $z$ is flat and sharply decreases towards zero near the free surface, i.e., it does not show a peak that would favor the formation of crystallites, we compute the average density profile of FK bonds during deposition and after relaxation (see Fig.~\ref{fig:rho_FK}). 
From it, we can notice the formation of a broad peak in $\langle\rho_{FK}\rangle$ with time corresponding to the formation of a critical nucleus (starting on average at $t>t_2$), as we verified by computing the average size of the main cluster with time (not shown).
Together with the $\langle\rho_{FK}\rangle$ for the DL, in Fig.~\ref{fig:rho_FK}, we show its behavior for the QL at two times, $t=t_r$ and $2t_r$. We can notice that after a time $t_r$, the QL shows a lower concentration of FK bonds compared to the DL relaxed for the same duration $t_r$, and the distribution in $z$ is asymmetric with a peak emerging near the free surface. This difference between the DL and QL is due to the higher stability reached by the DL during the deposition process compared to the QL. As shown in Ref.~\cite{leoni2023}, the internal potential energy and the specific volume in the DL are lower compared to the QL, while the concentration of LFS is correspondingly higher in a range of temperatures around $T_g$. Indeed, the symmetrical shape of $\langle\rho_{FK}(z)\rangle$ for the DL shown in Fig.~\ref{fig:rho_FK} is due to the fact that all the particles forming the DL were near the free surface at some point during the deposition process, while for the QL particles near the substrate never come close to the free surface during the quench process when $T<T_g$.
After a relaxation of $2t_r$, the peak in $\langle\rho_{FK}\rangle$ shown by the QL near the free surface overcomes the value of the DL (which has been relaxed for a time equal to $t_r$), while it still does not keep up with it near the substrate. 

\begin{figure}[t]
\begin{center}
\includegraphics[width=8.5cm]{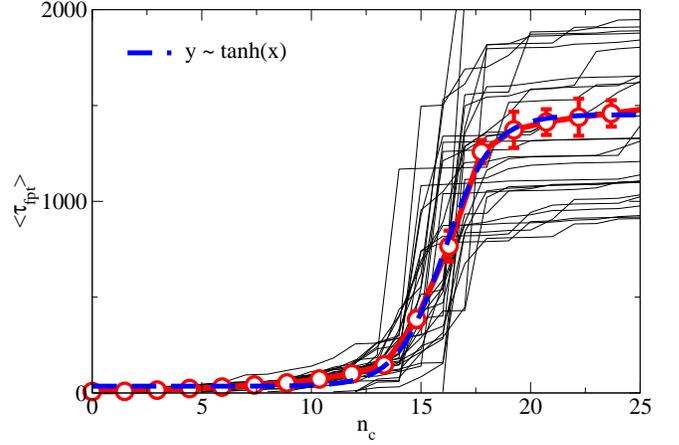}
\caption{\label{fig:tau_fpt} Mean first passage time $\langle\tau_{fpt}\rangle$ versus main cluster size $n_c$ for 30 seeds (black lines) and for the average (red circles) together with the hyperbolic tangent fit (blue dashed line)} at $T=0.4$.
\end{center}
\end{figure}

Usually, to estimate the size of the critical nucleus $n_c^*$, one can compute the mean first passage time $\langle\tau_{fpt}\rangle$ defined as the average time elapsed until the appearance of a nucleus of size $n_c$ in the system.
In Fig.~\ref{fig:tau_fpt}, we show $\langle\tau_{fpt}\rangle$ for every single realization of the system (black lines) together with the average curve (red circles). The mean first passage theory \cite{reguera2005,wedekind2007} would predict a critical nucleus corresponding to the inflection point of $\langle\tau_{fpt}\rangle$, which in our case gives $n_c^*\simeq 17$. 
A more refined version of the mean first passage theory, including space- and time-dependent diffusivity, could give a more precise estimation of $n_c^*$.

\begin{figure}[t]
\begin{center}
\hspace*{-0.1cm}\includegraphics[width=8.5cm]{Fig_8ab.eps}
\hspace*{0.1cm}\includegraphics[width=8.5cm]{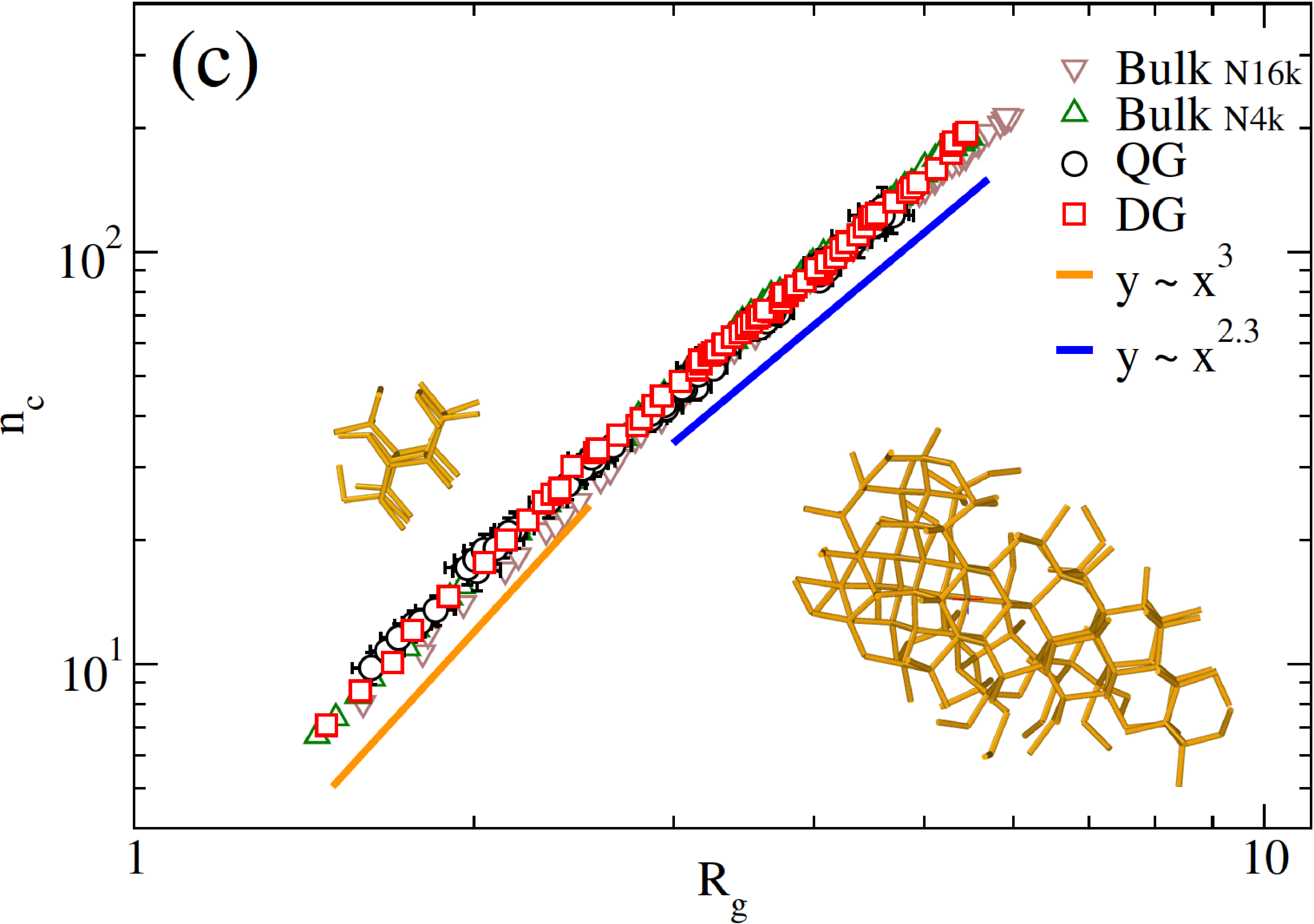}
\caption{\label{fig:moments} (a) average moments of inertia along $z$ ($I_{\perp}=I_{zz}$) and along a direction in the xy-plane ($I_{\|}=(I_{xx}+I_{yy})/2$) versus $n_c$ for the DL (red and blue squares, respectively) and QL (black and green circles, respectively). (b) Asphericity $A_s$ versus $n_c$.
(c) Nucleus size $n_c$ versus average radius of gyration $R_g$ for the DL (red squares), QL (black circles), and the bulk at two different system size (green upward-facing triangles for $N=4000$ and brown downward-facing triangles for $N=16000$). In the inset of panel (c) are shown typical nuclei of type 2 particles size 35 (left) and 250 (right) composed of a staking of two tetrahedral networks.}
\end{center}
\end{figure}
To investigate the structural properties of nuclei, we compute their moment of inertia with respect to the axis passing through their barycenter and parallel to one of the coordinate axis $x,y,z$ ($I_{xx}$, $I_{yy}$, $I_{zz}$, respectively). Due to the cylindrical symmetry of the system, we compute $I_{\perp}=I_{zz}$ and $I_{\|}=(I_{xx}+I_{yy})/2$. In panel (a) of Fig.~\ref{fig:moments}, we show that for both the DL and QL, the moment of inertia along different directions follows the typical $n_c^{5/3}$ scaling holding for nuclei which are symmetric on average. 
More specific information on the geometrical properties of nuclei can be obtained from the radius of gyration $R_g$ and the asphericity $A_s$ as a function of their size $n_c$.
We obtain $R_g$ and $A_s$ from the gyration tensor, so defined:
 \begin{equation}
S_{\alpha\beta}=\dfrac{1}{2n_c^2}\sum_{i=1}^{n_c}\sum_{j=1}^{n_c}(r_{\alpha}^{i}-r_{\alpha}^{j})(r_{\beta}^{i}-r_{\beta}^{j})     
\end{equation}
where $\alpha,\beta=x,y,z$, and $r_{\alpha}^i$ is the $\alpha$ component of the position vector of the particle $i$ belonging to the cluster.
The eigenvalues of $S_{\alpha\beta}$ are also called principal moments and can be written as the ordered elements $\lambda_x^2\leq\lambda_y^2\leq\lambda_z^2$.
The radius of gyration is defined as $R_g=\sqrt{Tr(S)}=\sqrt{\lambda_x^2+\lambda_y^2+\lambda_z^2}$, and the asphericity as $A_s=\lambda_z^2-(\lambda_x^2+\lambda_y^2)/2$.
The asphericity is $A_s=0$ for a perfect sphere and $A_s>0$ otherwise. The behavior of $A_s$ (Fig.~\ref{fig:moments}(b)) shows an increasing asymmetric character of nuclei with the increasing of their size, suggesting an asymmetric growth along the principal axis of inertia.
\begin{figure}[!t] 
\begin{center}
\includegraphics[width=8.5cm]{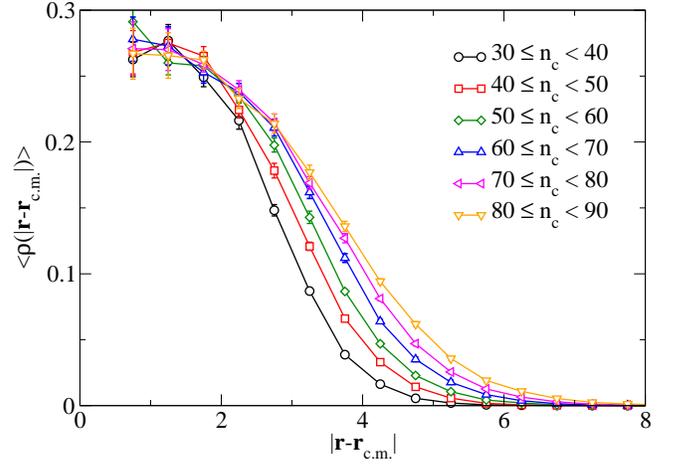}
\caption{\label{fig:nucleus} Average number density profile $\langle\rho(|{\bf r}-{\bf r}_{c.m.} |)\rangle$ of nuclei composed of type 2 particles forming FK bonds as a function of the distance from the center of mass $|{\bf r}-{\bf r}_{c.m.}|$ for different range of nucleus sizes $n$.}
\end{center}
\end{figure}
Indeed, from Fig.~\ref{fig:moments}(c) we can see that the size of a nucleus grows with $R_g$ as $n_c\sim R_g^3$ for small size, in agreement with the behavior of an isotropic 3D object, while it grows as $n_c\sim R_g^{2.3}$ for large nucleus size, suggesting either a sub-3D growth in this regime or a developing fractal or spongy morphology.

Considering all the results shown in Fig.~\ref{fig:moments}(a),(b),(c), together with snapshots inspection as those shown in the inset of panel (c), we conclude that nuclei grow following a sub-3D behavior after an initial 3D growth. 
Including type 1 particles in the cluster analysis would give further insights into the nuclei growth behavior with their size. Here we notice that the growth of nuclei showing multiple fully formed oblong tetrahedral networks (i.e., hexagonal rings of FK bonds forming a compact network) stacking onto each other (as for the right snapshot shown in panel (c) of Fig.~\ref{fig:moments}) suggests that even considering type 1 particles would give a sub-3D growth behavior and that the standard 3D growth could take over for larger clusters (i.e., for $n_c>300$).
During the formation of crystalline clusters, due to their $A_2B$ structural composition with A and B corresponding to
particles of type 1 and 2, respectively (as described in Sec.~\ref{sec:methods}), particles of type 1 are depleted. 
In Fig.~\ref{fig:moments}(c), alongside the results for the systems of size N=4000 particles, we also consider a larger bulk system ($N=16000$) to check whether this depletion affects the statistics of nuclei up to the size we considered in this work. From simulations of 20 independent trajectories, we find the same scaling laws for $n_c$ vs $R_g$ shown for both system sizes.
While the stability, and then the structure, of solid clusters can be size-dependent for small cluster size (as observed for systems as different as simple one-component liquids \cite{mossa2003} or a coarse-grained water model \cite{leoni2019}), the behavior of WAHN clusters for large cluster size is related to the preference of Laves phases such as the MgZn$_2$ to grow along the direction normal to the basal plane \cite{li2018}.

To further characterize the structure of nuclei, we computed their average radial number density profile $\langle\rho(|{\bf r}-{\bf r}_{c.m.}|)\rangle$ as a function of the distance from the center of mass $|{\bf r}-{\bf r}_{c.m.}|$ for different range of nucleus sizes (see Fig.~\ref{fig:nucleus}). 
Up to nuclei of size $\sim 40-50$, the density profile shows a 3D growth (profiles translate to larger values of $|{\bf r}-{\bf r}_{c.m.}|$ for increasing nucleus size $n_c$), while further increasing the nucleus size the density profile shows a sub-3D growth (profiles translate to larger values of $|{\bf r}-{\bf r}_{c.m.}|$ for increasing $n_c$ only in their tails).
This result confirms the change in the growth behavior of nuclei $n_c$ with $R_g$ for $n_c>40-50$ (see Fig.~\ref{fig:moments}). 
We notice that the scaling of the average moment of inertia of the nuclei with their size (right inset of Fig.~\ref{fig:moments}) is the same along the axis parallel and perpendicular to the interface, i.e. the nuclei do not have a preferential orientation on average. This means that the presence of a free surface does not affect the orientation of nuclei.

\section{Conclusion}
\label{sec:conclusion}

In conclusion, we investigated the nucleation process in a popular metallic glassformer (Wahnstr\"om mixture) during the layer formation by vapor deposition on a cold substrate and compared it with the conventional quenched solid in the same geometry and the bulk. We found that the as-deposited layer shows the significant enhancement of peculiar local ordered structures (FK bonds) compared to the QL (open red squares in Fig.~\ref{fig:FK}) and the bulk, even before relaxing the system (full symbols in Fig.~\ref{fig:FK}). Since the formation of these local structures is triggered by the enhanced mobility at the surface \cite{leoni2023}, the fact that all the particles of the deposited layer are part of the free surface at some point during the deposition process explains why the DL shows a larger content of FK bonds compared to the QL (see Fig.~\ref{fig:rho_FK}) and the bulk before the relaxation process takes place. 
On the other hand, we observe homogeneous nucleation in all the systems studied here and find that the nuclei share the same growth behavior, 3D-like up to a size of 40-50 type 2 particles and sub-3D for larger size due to the preference of clusters to grow along the basal plane.   
It is worth noting that it has recently been shown~\cite{hayton2023} that nucleation of water in film geometries remains bulk-like even for films size of the order of the critical nucleus. These findings provide further insights into nucleation behavior in confined environments for systems that are prone to crystallization, such as simple metallic glasses, and small molecular systems, such as water, for which nucleation during vapor deposition is commonly observed.

\vspace{0.5cm}

\begin{acknowledgments}
This work is dedicated to the 80th birthday of Myroslav Holovko, an outstanding scientist in the theory of the liquid state. We also wish to express solidarity with other Ukrainian scientists in these difficult times. 
F.L. and J.R. acknowledge support from the European Research Council Grant DLV-759187, partial support by ICSC – Centro Nazionale di Ricerca in High Performance Computing, Big Data and Quantum Computing, funded by European Union – NextGenerationEU, and CINECA-ISCRA for HPC resources.
H.T. acknowledges the grant-in-aid for Specially Promoted Research (JP20H05619) from the Japan Society of the Promotion of Science (JSPS).
\end{acknowledgments}

%

\end{document}